\documentclass[manuscript, natbib=false]{acmart}

\AtBeginDocument{%
  }

\acmBooktitle{FAccTRec '23: 5th FAccTRec Workshop: Responsible Recommendation,
Sept 18, 2023, Singapore}
\acmDOI{}

\usepackage[english]{babel}
\usepackage{amsthm}

\usepackage{array,booktabs,tabularx,multirow, multicol,colortbl,hhline,makecell}
\newcommand{\cell}[2]{\setlength{\tabcolsep}{0pt}\begin{tabular}{#1}#2 \end{tabular}}

\usepackage{graphicx,xcolor,caption,placeins,wrapfig}

\usepackage[capitalize]{cleveref}

\RequirePackage[
  datamodel=acmdatamodel,
  style=numeric,
  sorting=none,
  ]{biblatex}

\begin{document}

\title{Fairness Vs. Personalization: Towards Equity in Epistemic Utility}

\author{Jennifer Chien}
\email{jjchien@ucsd.edu}
\affiliation{%
  \institution{UC San Diego}
  \country{USA}
}

\author{David Danks}
\email{ddanks@ucsd.edu}
\affiliation{%
  \institution{UC San Diego}
  \country{USA}
}


\setcopyright{none}

\begin{abstract}
The applications of personalized recommender systems are rapidly expanding: encompassing social media, online shopping, search engine results, and more. These systems offer a more efficient way to navigate the vast array of items available. However, alongside this growth, there has been increased recognition of the potential for algorithmic systems to exhibit and perpetuate biases, risking unfairness in personalized domains. In this work, we explicate the inherent tension between personalization and conventional implementations of fairness. As an alternative, we propose equity to achieve fairness in the context of epistemic utility. We provide a mapping between goals and practical implementations and detail policy recommendations across key stakeholders to forge a path towards achieving fairness in personalized systems.
\end{abstract}

\keywords{fairness, personalization, recommender systems, equity, epistemic utility, epistemic harms, policy}

\maketitle

\section{Introduction}
Personalized algorithmic systems are being increasingly  deployed in a broad range of sectors. These algorithms hold the potential to provide more appropriate outputs on an individual basis by personalizing based on preferences, values, needs, or environmental conditions. Most prominently, recommender systems are now widely used to provide the most useful recommendations to each individual in the given domain\cite{Tufekci_2019}. 

At the same time, we have seen a significant increase in the recognition that algorithms can exhibit biases and produce unfair or unjust outcomes. There are many different potential sources of bias in algorithms and models, and hence many different responses may be appropriate or required. Algorithm development and deployment efforts now typically recognize the possibility of algorithmic bias, and the need to (often) do something to mitigate it\cite{Solmaz2020}.

We contend that these two desiderata for algorithms -- personalization and fairness -- stand in significant tension. At a high level, personalization is fundamentally about treating each individual in distinct ways; the goal is to \emph{not} give the same output for each individual, but rather to tailor the outputs to their specific situation. In contrast, fairness is fundamentally about treating individuals similarly; the algorithm ought to be ``the same'', in some sense, for everyone. Of course, this high-level gloss on the tension is far too quick: for example, fairness allows for differential treatment or outcomes, as long as it is based on morally or legally defensible grounds. Nonetheless, these high-level observations point towards a tension that, we contend, continues to hold when we look deeper. More specifically, we analyze fairness in the context of personalized systems using the notion of \emph{epistemic utility} -- essentially, the benefit that an agent receives by an epistemic improvement (e.g., reduction in uncertainty) -- and provide both practical and policy guidance about how to achieve fair, personalized systems.

\section{Preliminaries}
\label{Sec::Preliminaries}

\paragraph{Recommender System} We consider a recommender system as that which uses an algorithmic approach to provide an ordering or prioritization for items based on relevance to a user. This form of personalization is often based on their preferences, past behavior, or similarities with other users with the overarching goal of improving user experience and engagement.

\paragraph{Group Fairness}
Algorithmic fairness is typically framed as metric parity\footnote{within some tolerance as is legally permissible, e.g., the four-fifths rule\cite{watkins2022four}} across two or more relevantly comparable entities such that the difference between them is deemed socially, legally, and/or culturally irrelevant. For instance, group fairness may be defined as demographic parity, meaning equality in the likelihood of a given outcome conditioned on the group (e.g. gender, age, disability)\cite{dwork2012fairness}. Formally:
\begin{equation}
    P(\hat{y}~|~g=g_i) = P(\hat{y}~|~g=g_j) \forall g_i, g_j \in \{1, ..., G\}
\end{equation}

where $G$ is the total number of groups and $\hat{y}$ is the likelihood of a given outcome.

\paragraph{Epistemic Utility} We define epistemic utility as the value or usefulness of information in terms of improving knowledge, understanding, or beliefs. It encompasses the epistemic benefits gained from acquiring accurate, reliable, and relevant information. These benefits could, of course, enhance decision-making, prediction, problem-solving, or overall intellectual development, but epistemic utility refers only to the benefits in terms of the agent's epistemic state (even if those improvements do not immediately lead to improved outcomes). It thereby captures the intrinsic value of acquiring knowledge and understanding. It underscores the role of information as a valuable resource for intellectual growth, social mobility and production, and captures the motivational value of information for people's behavior.

\section{Related Work}
\label{Sec::Related Work}
This work is most closely related to efforts to understand fairness in recommender systems, many of which define fairness as parity in some performance measure across some sensitive attribute~\cite{mehrotra2018towards, sonboli2020opportunistic, liu2019personalized, boratto2021interplay, gomez2021winner, dash2021umpire}. Methods for individual notions of fairness similarly require a domain-specific similarity function and focus on trade-offs between individual and group fairness\cite{zhu2018fairness, edizel2020fairecsys}. In contrast, we start from the position that fairness requires equity across all individuals, rather than being defined solely through comparisons. This approach enables us to account for cases of unfairness where everyone is worse off (including situations with non-comparative injustices), not only those in which one group is.

More specifically, prior work has aimed to ensure proportional representation and exposure, often in e-commerce settings. Measures such as disparate exposure and visibility still require access to sensitive or group attributes~\cite{boratto2021interplay, gomez2021winner, dash2021umpire}. Calibration, in which the expected proportions of predicted classes match those observed in available data, implements group fairness without explicit knowledge of sensitive attribute status\cite{abdollahpouri2023calibrated, abdollahpouri2021user, wundervald2021cluster, da2021exploiting, deldjoo2023fairness}. However, its application in domains such as news can contribute to hegemonic regimes of representation, making it limited to applications where less diversity is valued by users\cite{deldjoo2023fairness}. Additionally, fairness definitions that target disparities in utility compare average sensitive attribute group performance rather than providing assurances for individuals or the entire population as a whole~\cite{zhu2018fairness, deldjoo2019recommender, deldjoo2020adversarial, li2021user, }. 


Concerns about unfairness in personalized algorithms is connected to research on epistemic injustice\cite{fricker2007epistemic}. The core concern in this latter work is that epistemic limitations may impair an agent's ability to recognize or respond to injustice. For example, if an individual does not have the concept of `racial discrimination', then they might fail to detect its occurrence despite suffering harms from it. That is, the injustice is not only about what happens to the individual, but also about their (epistemic) inability to recognize, understand, and respond to it. Personalized systems can readily lead to different individuals having different information -- in fact, that is one goal of such systems -- and so we expect that individuals would develop different concepts as a result. Hence, personalized systems can significantly raise the risks of epistemic injustice. 

Research on over-personalization and information access is similarly closely related. While the influence of personalization in filter bubbles and echo chambers has been denied for purely technical systems \cite{courtois2018challenging, dutton2019searching, nechushtai2019kind}, when considering human interactions, the influence becomes somewhat complex. This is, in part, due to significant variability in how users engage with a site, accept or reject information, and perceive the impartiality of the content they receive \cite{dillahunt2015detecting, ekstrom2022self, makhortykh2021can, teevan2008people, Shelton_2017}. Our research adopts a sociotechnical approach to define and scope the problem, taking into account the concept of epistemic utility and making reasonable assessments of human effort.

\section{Core tension: Personalization and standard approaches to fairness}
\label{Sec::Impossibility Argument}
Fairness is often defined as metric parity across similar groups or individuals, where determining similarity across units remains an open question. In search settings, one might start by grouping together individuals with similar values and interests, as, on fairness grounds, they may be expected to have a similar disambiguation or content recommendation experience. However, determining values can be difficult to operationalize because explicit and implicit preferences may be conflicting or difficult to collect~\cite{charlesworth2022patterns}. In addition, the level of granularity by which to measure such values remains unknown. In this section, we construct arguments for why user similarity of values is difficult, impractical, or perhaps even impossible to determine, thereby making similarity-based fairness definitions difficult to deploy or conceptually impossible.

\paragraph{Sensitive attributes are problematic and coarse proxies for values.}
The simplest approach to constructing ``similar'' groups for fairness considerations could use sensitive attributes (i.e. race, ethnicity, gender, age, religious affiliation, disability, etc). However, this approach assumes that individuals with the same sensitive attributes not only face similar struggles, lived experiences, discrimination, and prejudice, but also have relevantly similar values. In practice, this approach would likely result in the treatment of minority groups as monoliths -- perhaps even as a homogeneous `Other'. This is problematic not only in its instrumental failures of insufficient specificity and complexity, but also in its intrinsic treatment of a single sensitive attribute as representative of all values. Additionally, conditional groupings only ensure parity guarantees for the average member of a group, not for each individual nor necessarily an individual that exists, therefore enabling unfairness for anyone that doesn't satisfy such assumptions.

Taking an intersectional approach, i.e. considering subgroups of sensitive attributes may better approximate similarity of values. Consider, for example, race and gender, meaning groups such as ``Black Women'' and ``White Men''. This still considers an entire group of people homogeneously, thereby rendering the plight of, for example, all Black Women into a singular experience. In the extreme, considering all possible sub-groups across all sensitive attributes faces practical challenges, as each subgroup becomes impractically small (for data availability purposes), thereby reifying sensitive attributes in a constant struggle for relative popularity and public consciousness~\cite{kong2022intersectionally}.

\paragraph{User search history is noisy, underdetermines values, and can vary temporally.}
An alternative approach would be to use the individual's behavior to determine the relevant group for fairness considerations. For instance, in the context of a search engine, we could condition on exact search history. Unfortunately, this approach faces three open challenges which we highlight in the context of search, though the problems naturally generalize to other systems that use observed behavior to personalize on the basis of unobserved (inferred) features.

First, not all search behaviors are equally informative. For example, a search query may be made due to different situational demands (e.g., homework, third party question) or may arise from very different motivations (e.g., devil's advocate vs. searching for validative or supporting arguments). Hence, a specific history of search queries may or may not be informative about an individual's values; the history alone is not sufficient. 

Second, even if search behaviors are ``genuine,'' they are noisy indicators for the totality of an individual's interests or values. For example, the individual agent might not search for a topic they already have sufficient information on, but it may still be a core value. More generally, consider two users with identical search histories at any given point in time. Conditioning on search history would imply that they should receive similar experiences. However, it is highly unlikely that these individuals have the exact same sets of values and interests: that is, multiple sets of values and interests could be consistent with the observed behavior (i.e., the space of values are underdetermined by the evidence). Therefore, since queries subject to the limitations of observational data and missing contextual information, conditioning on search history alone has the potential to greatly under-specify an individual's values.

Finally, search history provides no guarantees of stability over time. Consider, again, two individuals with the same user history. Adding a different element to each user's search history could lead them to be placed in fundamentally different groups. More generally, at any point in time, there are infinitely many ways that a search history could progress, thereby making it impossible to guarantee fairness over time for these two individuals, let alone all. Fairness groupings not sensitive to the time and search history, therefore, may not ever converge to fair groupings in the limit. This echoes a similar problem in fairness research, where fairness constraints at each round of deployment do not ensure long-term aggregate parity in the same metric~\cite{hu2018short}. While this has not rendered particular fairness metrics obsolete, this instability points to ambiguity in the extent to which fairness must provide guarantees temporally.

\section{A Different Approach}
\label{Sec::New Definitions}

\subsection{Personalization}
Delving further into the notion of a system tailored to each person's needs, let us define personalization. Given a search engine, companies may strive to deliver a service that maximizes epistemic utility and convenience. In order to avoid wasting time, resources, or eroding consumer patience, search engine providers may optimize to disambiguate queries as efficiently as possible. Increases in efficiency can be made by leveraging all accessible information about a user. This may include search history, publicly available data, social media profiles, and inferred or disclosed demographic information to accurately disambiguate a user's query. This motivates the following definition:

\begin{definition}
    \textbf{Infinite Personalization.} Consider a platform that recommends relevant content to a user $u_j$ tailored to their previous history $h_{j, t}$ at time $t$. Since the space of all possible information is very large, for each piece of content $c_i\in C$, the system calculates probabilities $P(c_i ~|~ h_{j, t})$ estimating the likelihood of relevance. We define infinite personalization as the phenomena in which a system is allowed to produce extremely fine-grained personalization. Formally, there exists some $c_i$ such that $P(c_i ~|~ h_{j, t}) < \theta$ and $P(c_i ~|~ h_{j, t+\epsilon}) < \theta$, meaning that there exists some content such that it is arbitrarily improbable $< \theta$ to recommend a particular piece of content $c_i$ to a user, even if a user were to add $\epsilon$ queries to their search history.
    
    For instance, user $u_j$ with a search history exclusively composed of cats inputs ``bengal toys'' into a search engine and receives recommendations about Bengal cat toys. We consider the system to be exhibiting infinite personalization if the user can take $\epsilon$ queries about dogs (or anything else) and still not receive a recommendation for Bengal dog toys when searching ``bengal toys''.
\end{definition}

Having defined personalization and it's effects in the extreme case, we now define fairness as equity and provide motivation for the outcome we target.

\subsection{Equity}
In ensuring parity across similar groups, fairness guarantees are only made across those deemed sufficiently similar. An equity-based approach, however, centers around every individual achieving their desired outcome. Such approaches promote social cohesion, collective liberation, and address long-standing inequities, such as poverty, public health, social mobility, and overall well-being. Moreover, they tackle concerns regarding temporal stability and robustness by establishing a consistent criterion for each deployment phase. Equity-based approaches lend themselves to audits, legislation, and transparency, as they define a singular outcome requirement for all users. An open challenge, however, is what metrics to define equity with respect to, as each metric targets a different kind of unfairness. 

Equity across \emph{quality of information} targets unfairness when one person gets more utility from their query compared to a different query of another person. For example, this type of equity requires that we equalize the number of recommendations and quality of information provided for any given query. However, this is infeasible as we cannot control the quantity of information known for a given topic. Thus, given that there are infinitely many distinct queries any user can make, ensuring that each query has the same number of relevant articles would require perfect knowledge.

Equity in \emph{speed of access to information} targets unfairness when one person receives the relevant information faster than another. This equalizes the utility for every user by either: slowing down or acting intentionally uncooperatively for queries that are more easily disambiguated, or speeding up the service for ``slower'' queries. Both are infeasible. The former fails to maximize overall utility and sacrifices function in the name of fairness. This in turn renders adoption by stakeholders subject to capitalistic pressures even more unlikely. The latter is also infeasible, as we cannot directly control disambiguation rate and therefore speed up some queries over others. More generally, information is an infinitely shareable resource, so providing information to one person does not deprive someone else the benefit of that same information. Therefore speed of access to information is a poor target for an equity-based intervention.

In contrast, equity in \emph{epistemic utility} targets unfairness in personalized access to the information resources. This is motivated by the intuition that although we cannot directly control the availability of information or disambiguation rate, everyone should have some baseline access to all information. Thus, we propose an upper bound on the number of queries to access to any piece of information. This means that if the information does exist, all users should be able to access it by exerting a ``reasonable'' amount of effort. This does not require that all queries have the same accessibility or disambiguation, nor that all users be provided the exact same service. It does provide a soft guarantee that everyone is able to achieve some base-level epistemic utility without requiring the grouping or comparison of individuals on the basis of their values, needs, or histories. Epistemic utility, particularly through the internet, is recognized as a universal human right by Article 19 of the Universal Declaration of Human Rights~\cite{UNESCO_righttoinformation}, stating that, ``Everyone has the right to freedom...to seek, receive and impart information and ideas through any media and regardless of frontiers.''

\begin{definition}
    \textbf{$\epsilon$-Equity Fairness.} Given a user $u_j$ with history $h_{j, t}$, we define $\epsilon$-Equity fairness as the upper bound ($\epsilon$) of the number of additions to the search history such that for any given content $c_i\in C$, adding $\epsilon$ queries to the history will result in $P(c_i ~|~ h_{j, t+\epsilon}) \geq \theta$, enabling the user to obtain the relevant content. Here, we consider the case in which the relevant content does not exist as out of scope and focus purely on access.
\end{definition}

Embracing equity as a fundamental principle of fairness presents a range of complex challenges. These challenges encompass defining a baseline level of equity for all users, determining the pertinent factors that contribute to establishing such a baseline, and deliberating whether these considerations should be domain-specific or universally applicable. Rather than proposing a singular standard, we propose adopting a heuristic operationalization that prioritizes those who are most disadvantaged in terms of knowledge and understanding. Our intention is not to argue for a strictly prioritarian perspective, but that the incremental advancements can achieve the ultimate goal of equity in the long-term. Thus, we advocate for equity as a guiding concept to achieve equality by inductively reaching fairness in the limit.

Having established definitions infinite personalization and $\epsilon$-equity fairness, next we flesh out the tension between the two.

\section{Inherent Conflict and Trade-offs}
\label{Sec::Tradeoffs}

Infinite personalization, we posit, comes at a cost to $\epsilon$-equity fairness. More explicitly, in infinite personalization, the relevance probability of some information $c_i$ to a user can be less than $\theta$. This is due to the effects of personalization: indexing and re-ranking for efficient navigation across the vast space of information. However, this violates $\epsilon$-equity fairness, as some information will be effectively rendered unreachable within a fixed window of queries. We provide some examples of trade-offs below:

\begin{itemize}
    \item Search engine personalization prioritizes results relative to a user's input query and prior history. For any given topic orthogonal to a user's past history, personalization may rank desired items far beyond the average, expected, or typical number of results any given user explores, rendering it practically infeasible to reach. 
    \item Social media serves many functions: dissemination of news and information, entertainment, and social connection. Personalization in the extreme may contribute to propagation of mis- and disinformation, echo chambers, filter bubbles, radicalization, and social disconnection.
    \item Github Copilot (developed by OpenAI) focuses on code completion, providing suggestions for code lines or entire functions directly integrated into interactive development environments. Here, personalization has the potential to provide codebase-specific suggestions and completion, but may also infringe on a coder's ability construct novel functionality. 
    \item Google Bard is a language model for synthesizing insights for settings with a wide range of opinions, perspectives, or no right answer. As an augmentation to search-engine information retrieval, this may facilitate more efficient ordering and disambiguation of search results, but may also infringe on access to seemingly irrelevant information.
    \item ChatGPT is a general-purpose large language model (LLM) designed to engage in human-like conversations and answer a wide range of questions. Personalization in an extreme case, may lead to conversational loops -- or conversations that remain within the scope of a single topic, rather than having the flexibility to move between topics despite a user's prompt or request. \footnote{Here we list several examples that use LLMs due to their relevance to personalization, information synthesis and curation, and appeals to human-language usability. We acknowledge that there are many open relevant problems related to personalization, such as contribution or exacerbation of representational harms. For the scope of this work, we focus on how such systems can impact access to information.}
\end{itemize}
The key takeaway of this list is not that personalization necessitates unfairness or inequity, nor that fairness is not achievable in personalized systems. Rather, to  emphasize that personalization \emph{without consideration of fairness} can readily lead to systems that deny some individuals a base-level of expected epistemic utility.

\section{Policy Goals and Examples of Implementations}
\label{Sec::Policy}
While this conceptual analysis provides guidance about how best to think about fairness in the context of personalized algorithms, it does not provide guidance about how to achieve fairer, less biased systems. As we turn to more practical recommendations, we start by considering key stakeholders and their associated roles (\cref{Fig::Stakeholder Roles}). This creates a lexicon and clarifies the expected function of each stakeholder. 

\begin{table}[htbp]
\centering
\resizebox{\linewidth}{!}{
\begin{tabular}{p{0.2\linewidth}p{0.65\linewidth}}
  \toprule
  \textbf{Stakeholder} & \textbf{Role} \\
  \toprule
  \textbf{Government} & Develop guidelines, principles, procedures, regulations, and standards for safe, sustainable deployment. May hold the power to create structural supporting systems directly within the government or financially support external systems of enforcement of such regulations. May also enact other means of operationalizing policy, such as hosting interdisciplinary research summits or sending representatives to build collaborations with other stakeholders to better inform outputs. This stakeholder may also set roles and norms for other stakeholder responsibilities and consequences.\\
  \midrule
  \textbf{Civil Society} & Conduct evaluations and investigations of algorithms and software developed by others. This stakeholder may serve as an external, third-party entity for conducting unbiased evaluations of performance and compliance with policy. May be responsible for developing novel methods of measurement.\\
  \midrule
  \textbf{Industry} & Implement technical methods, tools, and technological innovations, usually manifesting as a consumer-facing service or product. May be subject to practical implementation constraints (i.e. financial, competition, or legislative). While a particular product may be theoretically feasible, it may not be practically viable or sustainable for a company to produce. Responsibilities may also include foreseeing and adequately mitigating harms of deployed products on those directly and indirectly affected. \\
  \midrule
  \cell{p{\linewidth}}{\textbf{Academia (broad)}} & Develop and implement technical, theoretical, and conceptual knowledge without direct consideration of profits. Results can include work that builds upon understanding, implications, or evaluation methods (i.e. downstream impacts, sociotechnical approaches). This stakeholder has the potential to act as an unbiased actor in developing best-practices for sustainable, long-term practices, potentially at the direct cost of profits (i.e. considering environmental, social, and cultural norms and influences).\\
  \midrule
  \textbf{General Public} & Provide input, feedback, criticism, and opinions on direction and alignment with values for research and innovation via implicit and explicit signals on the order of communities, groups, and/or individuals. Public support (or lack thereof) can be collected across a multitude of avenues and granularities, including financial support, interest, engagement, collective action, protests, organizations, and projects. \\
  \bottomrule
\end{tabular}
}
\caption{Stakeholder Roles. We define the expected roles of stakeholders for clarity of responsibilities enumerated in \cref{Fig::Policy Examples by Stakeholder}. We note that these responsibilities are not restrictive. There are many ways in which one stakeholder may overlap or even take over the responsibilities of another. For instance, if a company develops a policy for user privacy while it's competitors do not, self-imposed constraints on data collection may impact their product performance. However, as this design value receives public support, this can become a standard competitive service without the intervention of government or civil society.} 
\label{Fig::Stakeholder Roles}
\end{table}

Each of these stakeholders has a wide range of actions available that can lead to fairer personalized systems (see \cref{Fig::Policy Examples by Stakeholder}). These actions include a wide range of governance mechanisms, ranging from hard (e.g., regulation) to soft (e.g., social norms). We emphasize that there is no single response to ``fix'' unfairness in personalized systems: inevitably, many interdependent actions will likely need to be taken in order to make progress on this problem. Moreover, these are by no means intended to be exhaustive, but rather provide a grounded starting point that emphasizes the necessity of cross-stakeholder collaborations. 

\begin{table}[htbp]
\centering
\resizebox{\linewidth}{!}{
\begin{tabular}{p{0.22\linewidth}p{0.7\linewidth}}
  \toprule
  \textbf{Goal} & \textbf{Example Implementation/Operationalization}\\
  \toprule
  \cell{l}{\textbf{Mitigation}} &  
  \textbf{Industry, Academia:} Develop algorithms that facilitate connections between disjoint parts of the data manifold. This might be analogous to connecting different parts of the internet together through links, but done by creating arbitrary injections that fabricate similarities, covariances, or links between disjoint parts of the training data. \\
  & \textbf{Government:} Construct structural incentives (financial, prestige/recognition, standards) for achieving baseline levels of interconnectivity to ensure equitable access.\\
  \midrule
  \cell{l}{\textbf{Audit}} & 
  \textbf{Industry, Academia, Civil Society:} Develop measures that quantify epistemic utility for users using a range of behaviors and data (e.g. tool use, experienced frustration, time to complete a task, etc.). Produce formal, domain-specific representations of distributions of utility, such as non-uniform distributions being indicators of threats to fairness. Measure utility and study barriers to utility for non-users and those who have left the platform. \\
  & \textbf{Civil Society:} Report the utility distributions for non-users to compare against those that are already using the service. \\
  & \textbf{Academia:} Defining and measuring downstream impact may be essential to thorough quantification of disparities and unfairness. Downstream impacts such as social influences on platform usage can be influenced by an individual's experience. \\
  & \textbf{Government:} Regulation for timing and structure of fairness audits for personalized systems, including guidance on composition of internal and external stakeholders.\\
  \midrule
  \cell{l}{\textbf{Transparency}} & 
  \textbf{Industry, Civil Society:} Disclosure of aforementioned distributions of utility across various groups of users. Reports should include comments about whether these distributions are problematic, what are the proposed solutions, and on what timeline. \\
  & \textbf{Government:} Clear consequences and/or reparations for users affected throughout improvement and for continued negligence.\\
  & \textbf{General Public:} Active engagement and feedback on reports to inform appropriate consequences and informed public (dis)favor.\\
  \midrule
  \cell{l}{\textbf{Individual Control}} & \textbf{Industry, Academia:} Design control mechanisms over the degree of exploration to a query when serving recommendations. This may include prompting users for additional input before producing a prediction. Alternatively, allowing users to reset their history or set manual preferences over personalization filters.\\
  & \textbf{Government:} Regulation and enforcement of satisfaction and compliance.\\
  & \textbf{General Public:} Participation and feedback on norm setting of what features should be available, how accessible services should be, determine embedded values such as degree of ``reasonableness'' that are feasible and realistic for all users. \\
  \midrule
  \cell{l}{\textbf{Education \& Awareness}} & 
  \textbf{Industry, Academia, Government, General Public:} Workshops and training sessions can facilitate comprehension of the dangers, limitations, and better calibrate users to appropriate expectations of functionality. These workshops can include education on rights and how to advocate for them. For instance, if companies fail to comply, where can people file their grievances so they can be aggregated and collectively analyzed? \\
  & \textbf{Industry, Academia, Civil Society:} Development of interactive tools that enable visualization and comparison of an individual's utility compared to the general population or those relevantly comparable. \\
  & \textbf{Civil society, Government:} Regulation and continued measurement to ensure such tools flag anomalous data and ensure improvement within some time-frame. Create and publicly release of reports of corrections, adjustments, and investigations made to remedy grievances.\\
  \bottomrule
\end{tabular}
}
\caption{Example Policy Interventions and Goals by Stakeholder. Although we separate actions by stakeholders, we emphasize that these goals require collaborations and contributions across multiple stakeholders to achieve such solutions.}
\label{Fig::Policy Examples by Stakeholder}
\end{table}

\clearpage

\printbibliography

@book{fricker2007epistemic,
  title={Epistemic injustice: Power and the ethics of knowing},
  author={Fricker, Miranda},
  year={2007},
  publisher={Oxford University Press}
}

@article{teevan2008people,
  title={How people recall, recognize, and reuse search results},
  author={Teevan, Jaime},
  journal={ACM Transactions on Information Systems (TOIS)},
  volume={26},
  number={4},
  pages={1--27},
  year={2008},
  publisher={ACM New York, NY, USA}
}

@misc{Shelton_2017, title={Council post: The value of search results rankings}, url={https://www.forbes.com/sites/forbesagencycouncil/2017/10/30/the-value-of-search-results-rankings/}, journal={Forbes}, publisher={Forbes Magazine}, author={Shelton, Kelly}, year={2017}, month={Nov}}

@misc{UNESCO_righttoinformation, url={https://www.unesco.org/en/right-information}, journal={UNESCO.org}}

@article{charlesworth2022patterns,
  title={Patterns of implicit and explicit attitudes: IV. change and stability from 2007 to 2020},
  author={Charlesworth, Tessa ES and Banaji, Mahzarin R},
  journal={Psychological Science},
  volume={33},
  number={9},
  pages={1347--1371},
  year={2022},
  publisher={SAGE Publications Sage CA: Los Angeles, CA}
}

@article{watkins2022four,
  title={The four-fifths rule is not disparate impact: a woeful tale of epistemic trespassing in algorithmic fairness},
  author={Watkins, Elizabeth Anne and McKenna, Michael and Chen, Jiahao},
  journal={arXiv preprint arXiv:2202.09519},
  year={2022}
}

@inproceedings{dwork2012fairness,
  title={Fairness through awareness},
  author={Dwork, Cynthia and Hardt, Moritz and Pitassi, Toniann and Reingold, Omer and Zemel, Richard},
  booktitle={Proceedings of the 3rd innovations in theoretical computer science conference},
  pages={214--226},
  year={2012}
}

@inproceedings{kong2022intersectionally,
  title={Are “Intersectionally Fair” AI Algorithms Really Fair to Women of Color? A Philosophical Analysis},
  author={Kong, Youjin},
  booktitle={2022 ACM Conference on Fairness, Accountability, and Transparency},
  pages={485--494},
  year={2022}
}

@inproceedings{hu2018short,
  title={A short-term intervention for long-term fairness in the labor market},
  author={Hu, Lily and Chen, Yiling},
  booktitle={Proceedings of the 2018 World Wide Web Conference},
  pages={1389--1398},
  year={2018}
}

@article{boratto2021interplay,
  title={Interplay between upsampling and regularization for provider fairness in recommender systems},
  author={Boratto, Ludovico and Fenu, Gianni and Marras, Mirko},
  journal={User Modeling and User-Adapted Interaction},
  volume={31},
  number={3},
  pages={421--455},
  year={2021},
  publisher={Springer}
}

@misc{Tufekci_2019, title={How recommendation algorithms run the world}, url={https://www.wired.com/story/how-recommendation-algorithms-run-the-world/}, journal={Wired}, publisher={Conde Nast}, author={Tufekci, Zeynep}, year={2019}, month={Apr}}

@inproceedings{abdollahpouri2021user,
  title={User-centered evaluation of popularity bias in recommender systems},
  author={Abdollahpouri, Himan and Mansoury, Masoud and Burke, Robin and Mobasher, Bamshad and Malthouse, Edward},
  booktitle={Proceedings of the 29th ACM Conference on User Modeling, Adaptation and Personalization},
  pages={119--129},
  year={2021}
}

@misc{Solmaz2020, title={Why algorithmic fairness?}, url={https://selects.acm.org/selections/why-algorithmic-fairness}, journal={ACM Selects}, publisher={ACM}, author={Solmaz, Gurkan and Tithi, Jesmin Jahan and Miguel de Joya, Juan}, year={2020}, month={Oct}}

@article{makhortykh2021can,
  title={Can filter bubbles protect information freedom? Discussions of algorithmic news recommenders in Eastern Europe},
  author={Makhortykh, Mykola and Wijermars, Mari{\"e}lle},
  journal={Digital Journalism},
  pages={1--25},
  year={2021},
  publisher={Taylor \& Francis}
}

@article{nechushtai2019kind,
  title={What kind of news gatekeepers do we want machines to be? Filter bubbles, fragmentation, and the normative dimensions of algorithmic recommendations},
  author={Nechushtai, Efrat and Lewis, Seth C},
  journal={Computers in human behavior},
  volume={90},
  pages={298--307},
  year={2019},
  publisher={Elsevier}
}

@article{ekstrom2022self,
  title={Self-imposed filter bubbles: Selective attention and exposure in online search},
  author={Ekstr{\"o}m, Axel G and Niehorster, Diederick C and Olsson, Erik J},
  journal={Computers in Human Behavior Reports},
  volume={7},
  pages={100226},
  year={2022},
  publisher={Elsevier}
}

@article{dutton2019searching,
  title={Searching through filter bubbles, echo chambers},
  author={Dutton, William H and Reisdorf, Bianca C and Blank, Grant and Dubois, Elizabeth},
  journal={Society and the internet: How networks of information and communication are changing our lives},
  pages={228},
  year={2019}
}

@article{courtois2018challenging,
  title={Challenging Google Search filter bubbles in social and political information: Disconforming evidence from a digital methods case study},
  author={Courtois, C{\'e}dric and Slechten, Laura and Coenen, Lennert},
  journal={Telematics and Informatics},
  volume={35},
  number={7},
  pages={2006--2015},
  year={2018},
  publisher={Elsevier}
}

@inproceedings{dillahunt2015detecting,
  title={Detecting and visualizing filter bubbles in Google and Bing},
  author={Dillahunt, Tawanna R and Brooks, Christopher A and Gulati, Samarth},
  booktitle={Proceedings of the 33rd Annual ACM Conference Extended Abstracts on Human Factors in Computing Systems},
  pages={1851--1856},
  year={2015}
}

@article{edizel2020fairecsys,
  title={FaiRecSys: mitigating algorithmic bias in recommender systems},
  author={Edizel, Bora and Bonchi, Francesco and Hajian, Sara and Panisson, Andr{\'e} and Tassa, Tamir},
  journal={International Journal of Data Science and Analytics},
  volume={9},
  pages={197--213},
  year={2020},
  publisher={Springer}
}

@article{wundervald2021cluster,
  title={Cluster-based quotas for fairness improvements in music recommendation systems},
  author={Wundervald, Bruna},
  journal={International Journal of Multimedia Information Retrieval},
  volume={10},
  number={1},
  pages={25--32},
  year={2021},
  publisher={Springer}
}

@inproceedings{abdollahpouri2023calibrated,
  title={Calibrated recommendations as a minimum-cost flow problem},
  author={Abdollahpouri, Himan and Nazari, Zahra and Gain, Alex and Gibson, Clay and Dimakopoulou, Maria and Anderton, Jesse and Carterette, Benjamin and Lalmas, Mounia and Jebara, Tony},
  booktitle={Proceedings of the Sixteenth ACM International Conference on Web Search and Data Mining},
  pages={571--579},
  year={2023}
}

@inproceedings{zhu2018fairness,
  title={Fairness-aware tensor-based recommendation},
  author={Zhu, Ziwei and Hu, Xia and Caverlee, James},
  booktitle={Proceedings of the 27th ACM international conference on information and knowledge management},
  pages={1153--1162},
  year={2018}
}

@inproceedings{li2021user,
  title={User-oriented fairness in recommendation},
  author={Li, Yunqi and Chen, Hanxiong and Fu, Zuohui and Ge, Yingqiang and Zhang, Yongfeng},
  booktitle={Proceedings of the Web Conference 2021},
  pages={624--632},
  year={2021}
}

@inproceedings{deldjoo2020adversarial,
  title={Adversarial machine learning in recommender systems (aml-recsys)},
  author={Deldjoo, Yashar and Di Noia, Tommaso and Merra, Felice Antonio},
  booktitle={Proceedings of the 13th International Conference on Web Search and Data Mining},
  pages={869--872},
  year={2020}
}

@article{deldjoo2019recommender,
  title={Recommender systems fairness evaluation via generalized cross entropy},
  author={Deldjoo, Yashar and Anelli, Vito Walter and Zamani, Hamed and Bellog{\'\i}n, Alejandro and Di Noia, Tommaso},
  journal={arXiv preprint arXiv:1908.06708},
  year={2019}
}

@inproceedings{dash2021umpire,
  title={When the umpire is also a player: Bias in private label product recommendations on e-commerce marketplaces},
  author={Dash, Abhisek and Chakraborty, Abhijnan and Ghosh, Saptarshi and Mukherjee, Animesh and Gummadi, Krishna P},
  booktitle={Proceedings of the 2021 ACM Conference on Fairness, Accountability, and Transparency},
  pages={873--884},
  year={2021}
}

@inproceedings{gomez2021winner,
  title={The winner takes it all: geographic imbalance and provider (un) fairness in educational recommender systems},
  author={G{\'o}mez, Elizabeth and Shui Zhang, Carlos and Boratto, Ludovico and Salam{\'o}, Maria and Marras, Mirko},
  booktitle={Proceedings of the 44th International ACM SIGIR Conference on Research and Development in Information Retrieval},
  pages={1808--1812},
  year={2021}
}

@inproceedings{mehrotra2018towards,
  title={Towards a fair marketplace: Counterfactual evaluation of the trade-off between relevance, fairness \& satisfaction in recommendation systems},
  author={Mehrotra, Rishabh and McInerney, James and Bouchard, Hugues and Lalmas, Mounia and Diaz, Fernando},
  booktitle={Proceedings of the 27th acm international conference on information and knowledge management},
  pages={2243--2251},
  year={2018}
}

@article{deldjoo2023fairness,
  title={Fairness in recommender systems: research landscape and future directions},
  author={Deldjoo, Yashar and Jannach, Dietmar and Bellogin, Alejandro and Difonzo, Alessandro and Zanzonelli, Dario},
  journal={User Modeling and User-Adapted Interaction},
  pages={1--50},
  year={2023},
  publisher={Springer}
}

@article{da2021exploiting,
  title={Exploiting personalized calibration and metrics for fairness recommendation},
  author={da Silva, Diego Corr{\^e}a and Manzato, Marcelo Garcia and Dur{\~a}o, Frederico Ara{\'u}jo},
  journal={Expert Systems with Applications},
  volume={181},
  pages={115112},
  year={2021},
  publisher={Elsevier}
}

@inproceedings{liu2019personalized,
  title={Personalized fairness-aware re-ranking for microlending},
  author={Liu, Weiwen and Guo, Jun and Sonboli, Nasim and Burke, Robin and Zhang, Shengyu},
  booktitle={Proceedings of the 13th ACM Conference on Recommender Systems},
  pages={467--471},
  year={2019}
}

@inproceedings{sonboli2020opportunistic,
  title={Opportunistic multi-aspect fairness through personalized re-ranking},
  author={Sonboli, Nasim and Eskandanian, Farzad and Burke, Robin and Liu, Weiwen and Mobasher, Bamshad},
  booktitle={Proceedings of the 28th ACM Conference on User Modeling, Adaptation and Personalization},
  pages={239--247},
  year={2020}
}

\end{document}